\documentstyle[multicol,aps,prl,psfig]{revtex} 

\renewcommand{\narrowtext}{\begin{multicols}{2} \global\columnwidth20.5pc}
\renewcommand{\widetext}{\end{multicols} \global\columnwidth42.5pc}
\def\top#1{\vskip #1\begin{picture}(290,80)(80,500)\thinlines \put(
65,500){\line( 1, 0){255}}\put(320,500){\line( 0, 1){
5}}\end{picture}}
\def\bottom#1{\vskip #1\begin{picture}(290,80)(80,500)\thinlines \put(
330,500){\line( 1, 0){255}}\put(330,500){\line( 0, -1){
5}}\end{picture}}

\def\al{\alpha}
\def\be{\beta}
\def\ga{\gamma}
\def\de{\delta}
\def\ep{\epsilon}

\def\et{\eta}
\def\th{\theta}

\def\la{\lambda}

\def\ta{\tau}

\def\ph{\phi}

\def\ch{\chi}
\def\ps{\psi}

\def\Ga{\Gamma}
\def\De{\Delta}

\def\La{\Lambda}

\def\Up{\Upsilon}

\def\Ps{\Psi}
\def\Om{\Omega}

\def\fr#1#2{{{#1} \over {#2}}}

\def\prt{\partial}

\def\ket#1{|{#1}\rangle}

\def\half{{\textstyle{1\over 2}}}
\def\lsim{\mathrel{\rlap{\lower4pt\hbox{\hskip1pt$\sim$}}
    \raise1pt\hbox{$<$}}}
\def\gsim{\mathrel{\rlap{\lower4pt\hbox{\hskip1pt$\sim$}}
    \raise1pt\hbox{$>$}}}

\def\X{\hat X}
\def\Y{\hat Y}
\def\Z{\hat Z}
\def\x{\hat x}
\def\y{\hat y}
\def\z{\hat z}
\def\Re{\hbox{Re}\,}
\def\Im{\hbox{Im}\,}

\newcommand{\beq}{\begin{equation}}
\newcommand{\eeq}{\end{equation}}
\newcommand{\bea}{\begin{eqnarray}}
\newcommand{\eea}{\end{eqnarray}}
\newcommand{\rf}[1]{(\ref{#1})}

\begin{document}

\title{
Signals for CPT and Lorentz Violation in Neutral-Meson Oscillations
}    
\author{V.\ Alan Kosteleck\'y}
\address{Physics Department, Indiana University, 
          Bloomington, IN 47405, U.S.A.}
\date{IUHET 406, May 1999; accepted for publication in Physical Review D} 
\maketitle

\begin{abstract}
Experimental signals for indirect CPT violation 
in the neutral-meson systems are studied
in the context of a general CPT- and Lorentz-violating 
standard-model extension.
In this explicit theory,
some CPT observables depend on the meson momentum 
and exhibit diurnal variations.
The consequences for CPT tests 
vary significantly with the specific experimental scenario.
The wide range of possible effects is illustrated
for two types of CPT experiment presently underway,
one involving boosted uncorrelated kaons
and the other involving unboosted correlated kaon pairs. 
\end{abstract}


\narrowtext

\section{Introduction}
\label{intro}

The notion of studying CPT symmetry to high precision
by using the interferometric sensitivity of neutral-meson oscillations
to compare properties of mesons and their antiparticles 
dates from several decades ago \cite{lw}.
In recent years,
experiments with neutral kaons
have constrained the CPT figure of merit 
$r_K \equiv |m_K - m_{\overline{K}}|/m_K$
to about one part in $10^{18}$.
The most recent published result is
$r_K < 1.3 \times 10^{-18}$
at the 90\% confidence level 
from the experiment E773 at FNAL 
\cite{e773}.
A preliminary result corresponding to 
a value of $r_K$ below one part in $10^{18}$
has been announced by the CPLEAR collaboration
at CERN
\cite{cplear}.
Experiments now underway such as KTeV \cite{ktev}
at FNAL 
or KLOE \cite{kloe} 
at Frascati,
as well as other future possibilities \cite{gt},
are likely to improve these bounds.
High-precision CPT tests have also been performed 
with the neutral-$B$ system
by the OPAL 
\cite{opal}
and DELPHI 
\cite{delphi}
collaborations at CERN,
and other CPT measurements in the $D$, $B_d$, and $B_s$ systems 
are likely to be feasible 
in experiments at the various charm and $B$ factories.

On the theoretical front,
a purely phenomenological treatment of CPT violation
in the neutral-kaon system has also existed for decades
\cite{lw}.
It allows for indirect CPT violation 
by adding a complex parameter 
in the standard expressions relating the physical meson states
to the strong-interaction eigenstates.

Although a purely phenomenological treatment of this type 
is necessary in the absence 
of a convincing framework for CPT violation,
it is unsatisfactory from the theoretical perspective. 
Ideally,
CPT violation should be studied 
within a plausible theoretical framework
allowing its existence 
\cite{cpt98}.
The purely phenomenological treatment of CPT violation 
can be contrasted with the situation 
for conventional CP violation,
where a nonzero value  
of the phenomenological parameter $\ep_P$ 
for T violation
can be understood in the context of
the usual standard model of particle physics
\cite{rs,cj,ww}.

Over the past decade,
a promising theoretical possibility for CPT violation
has been developed.
It is based on spontaneous breaking of CPT and Lorentz symmetry
in an underlying theory
\cite{kps},
perhaps at the Planck scale
where one might plausibly expect modifications
to the conventional theoretical framework
arising from string theory 
or some other quantum theory of gravity.
It appears to be compatible both with experimental constraints
and with established quantum field theory,
and it leads to a general standard-model extension 
preserving gauge invariance and renormalizability 
that can be used as a basis for the phenomenology 
of CPT and Lorentz violation 
\cite{kp,cksm}.
The possibility therefore exists of using CPT tests 
as a quantitative probe of nature at the Planck scale.

An important feature of the standard-model extension
is its applicability to a broad class of experiments.
It provides a quantitative microscopic framework
for CPT and Lorentz violation
at the level of the standard model,
allowing the prediction of signals 
and the comparison and evaluation of experimental bounds.
Investigations related to this theory have 
to date been conducted for 
neutral-meson systems 
\cite{opal,delphi,kp,ckpv,ak},
tests of QED in Penning traps
\cite{pennexpts,bkr,gg,hd,rm},
photon birefringence and radiative QED effects
\cite{cksm,cfj,jk,pv},
hydrogen and antihydrogen spectroscopy
\cite{antih,bkr2},
clock-comparison experiments
\cite{ccexpt,kl},
muon properties
\cite{bkl},
cosmic-ray and neutrino tests
\cite{cg},
and baryogenesis
\cite{bckp}.

The strength of the CPT theorem
\cite{rs}
makes it difficult to create a theoretical framework 
for CPT violation that is plausible
and avoids radical revisions of established quantum field theory
\cite{cpt98,qmviol}.
It is therefore unsurprising that
in the context of the standard-model extension
the phenomenological parameters for CPT violation
turn out to have properties beyond those normally assumed.
Indeed,
it has been shown
that previously unexpected effects appear,
including momentum dependence 
(in both magnitude and orientation)
of the experimental observables for CPT violation.
The consequences for experimental signals can be substantial
and include, for example,
the possibility that diurnal variations 
exist in observable quantities
\cite{ak}.

The present work investigates the theoretical underpinnings
of experimental signals for CPT violation
in the context of the general standard-model extension.
One goal is to illustrate
how disparate the implications
of the momentum and time dependence can be
for different experimental scenarios
and thereby to encourage the detailed realistic simulations
and data analyses required for a satisfactory extraction
of CPT bounds from any specific experiment.

Some theoretical considerations applicable to all 
relevant neutral-meson systems
are presented in section \ref{theory},
along with some results specific to the kaon system.
Experiments on CP violation using neutral mesons
can be classified according to whether the mesons involved
are heavily boosted or not and whether they appear 
as uncorrelated events or as correlated pairs.
The possibility of CPT-violating effects dependent
on the momentum magnitude or orientation 
implies substantially different sensitivities to CPT violation
for these various classes of experiment.
In section \ref{experiment},
some consequences are developed for two illustrative cases.
One is exemplified by 
the KTeV experiment \cite{ktev} at FNAL,
in which a collimated beam of highly boosted uncorrelated mesons
is studied.
The other is exemplified by the KLOE experiment 
\cite{kloe}
at DAPHNE in Frascati,
which involves correlated meson pairs 
with a wide angular distribution at relatively low boost.
The results obtained provide intuition about
effects to be expected in various types of experiment,
including ones involving other neutral-meson systems.

\section{Theory}
\label{theory}

This section begins with 
some theoretical considerations
about CPT violation and its description
in the context of the general standard-model extension,
applicable to any of the four relevant types of neutral meson.
In what follows,
the strong-interaction eigenstates are denoted
generically by $P^0$,
where $P^0$ is one of $K^0$, $D^0$, $B_d^0$, $B_s^0$.
The corresponding opposite-flavor antiparticle 
is denoted $\overline{P^0}$.

A general neutral-meson state is a linear combination
of the Schr\"odinger wave function for $P^0$ and $\overline{P^0}$,
which can be represented by a two-component object $\Ps$.
The time evolution of the state is determined
by an equation of the Schr\"odinger form
\cite{lw}:
\beq
i\prt_t \Ps = \La \Ps
\quad ,
\label{seq}
\eeq
where
$\La$ is a 2$\times$2 effective hamiltonian.
The eigenstates of this hamiltonian represent the
physical propagating states and are generically denoted
$P_S$ and $P_L$.
The corresponding eigenvalues 
are 
\beq
\la_S \equiv m_S - \half i \ga_S
\quad , \quad 
\la_L \equiv m_L - \half i \ga_L
\quad ,
\label{mga}
\eeq
where $m_S$, $m_L$ are the propagating masses
and $\ga_S$, $\ga_L$ are the associated decay rates.
For simplicity,
here and in what follows
the subscripts $P$ are suppressed
on all these quantities and on the
components of the effective hamiltonian $\La$.

Flavor oscillations between $P^0$ and $\overline{P^0}$
are governed by the off-diagonal components of $\La$.
It can be shown 
\cite{lw}
that this system exhibits indirect CPT violation
\cite{fn0}
if and only if the difference of diagonal elements of $\La$
is nonzero, $\La_{11} - \La_{22} \neq 0$.
The effective hamiltonian $\La$ can be written
as $\La \equiv M - \half i \Ga$,
where $M$ and $\Ga$ are hermitian 2$\times$2 matrices
called the mass and decay matrices, respectively.
The condition for CPT violation therefore 
can in general be written
\beq
\De M - \half i \De \Ga \neq 0
\quad , 
\label{cptv}
\eeq
where
$\De M \equiv M_{11} - M_{22}$
and $\De \Ga \equiv \Ga_{11} - \Ga_{22}$.
Note that the elements $M_{11}$, $M_{22}$, $\Ga_{11}$, $\Ga_{22}$
are real by definition.

In the context of the general standard-model extension,
the dominant CPT-violating contributions to $\La$ 
can be obtained in perturbation theory,
arising as expectation values of interaction terms
in the standard-model hamiltonian
\cite{kps}.
The appropriate states for the expectation values
are the wave functions for 
the $P^0$ and $\overline{P^0}$ mesons
in the absence of CPT violation.
Since the perturbing hamiltonian is hermitian,
the leading-order contributions to the diagonal terms of $\La$
are necessarily real,
which means $\De\Ga = 0$.
The dominant CPT signal therefore necessarily resides
only in the difference of diagonal elements 
of the mass matrix $M$.
For the general standard-model extension,
it follows that the figure of merit 
\beq
r_P \equiv \fr{|m_P - m_{\overline{P}}|}{m_P}
\equiv \fr{|\De M|}{m_P}
\label{rp}
\eeq
provides a complete description 
of the magnitude of the dominant CPT-violating effects.
This can be contrasted 
with the usual phenomenological description,
for which the effects of $\De\Ga \neq 0$
should also be considered.
For example,
contributions involving the diagonal elements of $\Ga$
could in principle play an important role.

To make further progress,
it is useful to have an explicit expression 
for $\De M$ in terms of quantities appearing
in the standard-model extension.
This has been obtained in Refs.\ \cite{kp,ak}.
Several factors combine to 
yield a surprisingly simple expression for $\De M$.
Since the eigenstates of both the strong interactions
and the effective hamiltonian $\La$
are eigenstates of the parity operator,
and since charge conjugation is violated
by the flavor mixing,
any CPT-violating effects must arise from terms 
in the standard-model extension
that violate C while preserving P.
Also,
only contributions linear
in the parameters for CPT violation are of interest
because all such parameters are expected to be minuscule.
Moreover,
flavor-nondiagonal CPT-violating effects
can be neglected since they are suppressed relative
to flavor-diagonal ones.

The result of the derivation is 
\beq
\De M \approx \be^\mu \De a_\mu
\quad .
\label{dem}
\eeq
In this expression,
$\be^\mu$ is the four-velocity of the
meson state in the observer frame:
$\be^\mu = \ga (1, \vec \be )$.
Also, 
$\De a_\mu = r_{q_1}a^{q_1}_\mu - r_{q_2}a^{q_2}_\mu$,
where $a^{q_1}_\mu$, $a^{q_2}_\mu$
are CPT- and Lorentz-violating coupling constants
for the two valence quarks in the $P^0$ meson,
and where the factors $r_{q_1}$ and $r_{q_1}$
allow for quark-binding or other normalization effects 
\cite{kp}.
The coupling constants 
$a^{q_1}_\mu$, $a^{q_2}_\mu$
have dimensions of mass
and are associated with terms 
in the standard-model extension of the form 
$- a^q_\mu \overline{q} \ga^\mu q$,
where $q$ is a quark field of a specific flavor
\cite{fn1}.
It is perhaps worth noting that the flavor-changing experiments
on neutral mesons discussed here 
are the only tests identified to date that 
are sensitive to the parameters $a^q_{\mu}$,
so bounds from these experiments are of interest 
independently of any other tests of CPT and Lorentz symmetry.
Note also that 
the dependence of $\De M$ on the meson four-velocity
and hence on the meson four-momentum
is difficult to anticipate in the context of 
the usual purely phenomenological description of CPT violation,
where it seems reasonable
\it a priori \rm to take $\De M$
as independent of momentum.
However,
the momentum dependence is compatible 
with the significant changes expected 
in the conventional picture
if the CPT theorem is to be violated.

In the next section,
a few experimental consequences of the 
dependence on the meson four-momentum
magnitude and orientation are considered,
and illustrations of these consequences
for specific experiments are given.
These examples involve kaons,
for which a widely used variable for CPT violation
is denoted $\de_K$ \cite{lw}.
It can be introduced through the exact relation
between the eigenstates of the strong interaction
and those of the effective hamiltonian:
\bea
\ket{K_S} &=&
\fr{ (1 + \ep_K + \de_K) \ket{K^0}
    +(1 - \ep_K - \de_K) \ket{\overline{K^0}}  }
   { \sqrt{2( 1 + |\ep_K + \de_K|^2)}  }
\quad ,
\nonumber \\
\ket{K_L} &=&
\fr{ (1 + \ep_K - \de_K) \ket{K^0}
    -(1 - \ep_K + \de_K) \ket{\overline{K^0}}  }
   { \sqrt{2( 1 + |\ep_K - \de_K|^2)}  }
\quad .
\label{epde}
\eea
In the kaon system,
indirect T violation is small and
any CPT violation must also be small. 
This ensures that $\de_K$ is effectively 
a phase-independent quantity. 
However, $\ep_K$ does vary with the choice of phase convention.

Under the assumption that CPT and T violation
are both small,
$\de_K$ can in general be expressed as 
\beq 
\de_K \approx {\De \La}/{2\De \la}
\quad ,
\label{dk}
\eeq
where $\De \la$ is the difference of the eigenvalues
\rf{mga} of $\La$.
In terms of the mass and decay-rate differences
$\De m \equiv m_L - m_S$
and $\De \ga \equiv \ga_S - \ga_L$,
one has
\bea
\De \la &\equiv & \la_S - \la_L 
= - \De m - \half i \De \ga 
\nonumber \\
&=& - i \fr{\De m} {\sin\hat\ph} e^{-i\hat\ph}
\quad .
\label{dela}
\eea
In this expression,
$\hat\ph \equiv \tan^{-1}(2\De m/\De \ga )$
is sometimes called the superweak angle.
Note that a subscript $K$ is understood 
on all the above quantities.

In the context of the standard-model extension,
$\De \La = \De M$ is given by Eq.\ \rf{dem}.
For a meson with velocity $\vec\be$
and corresponding boost factor $\ga$,
Eqs.\ \rf{dem}, \rf{dk}, and \rf{dela} imply
\beq
\de_K \approx i \sin\hat\ph ~ e^{i\hat\ph} 
\ga(\De a_0 - \vec \be \cdot \De \vec a) /\De m
\quad .
\label{dek}
\eeq
The figure of merit $r_K$ in Eq.\ \rf{rp} becomes
\bea
r_K &\equiv & \fr {|m_K - m_{\overline{K}}|}{m_K}
\approx 
\fr{2 \De m} {m_K \sin\hat\ph} |\de_K|
\nonumber \\
&\approx & \fr { |\be^\mu\De a_\mu| }{m_K}
\quad .
\label{rk}
\eea
Using the known experimental values
\cite{pdg}
for $\De m$, $m_K$, and $\sin\hat\ph$
gives
\beq
r_K \simeq
2\times 10^{-14} |\de_K|
\simeq 2 \left| \be^\mu \fr {\De a_\mu} {1~ \rm GeV} \right|
\quad .
\label{rknum}
\eeq
A bound on $|\de_K|$ of about $10^{-4}$
therefore corresponds to a constraint on $|\be^\mu\De a_\mu|$
of about $10^{-18}$ GeV.

In the above expressions,
the explicit momentum dependence
arises from the dependence of $\De M$ on $\be_\mu$.
Since the eigenfunctions and eigenvalues of $\La$
depend on $M_{11}$ and $M_{22}$,
the possibility exists that 
there is also hidden momentum dependence in
the parameter $\ep_K$,
in the masses and decay rates
$m_S$, $m_L$, $\ga_S$, $\ga_L$,
and in the associated quantities 
$\De m$, $\De \ga$, $\hat\ph$.
However, 
all such dependence is suppressed
relative to that explicitly displayed above
because the CPT-violating contribution to
$M_{22}$ is the negative of the contribution to $M_{11}$. 
The sole linearly independent source 
of variation with momentum is therefore 
the difference $\De M$,
and only the parameter for CPT violation $\de_K$ 
is sensitive to $\De M$ at leading order. 
For example,
$\ep_K$ depends on $\De M$ at most 
through correction factors involving the 
square of the ratio $\De M/\De\la$,
which is of order $\de_K^2$,
so for all practical purposes 
conventional indirect CP (T) violation 
displays no momentum dependence in 
the present framework
\cite{bell}.
The same is true of the other quantities,
basically because a small difference between diagonal elements
of a matrix changes its eigenvalues only by amounts
proportional to the square of that difference.

\section{Experiment}
\label{experiment}

The implications of the four-velocity and
hence momentum dependence in the parameters 
for CPT violation can be substantial
for experiments with $P^0$ mesons.
The possible effects include 
ones arising from the dependence 
on the magnitude of the momentum
and ones arising from the variation 
with orientation of the meson boost
\cite{ak}.
Consequences of the dependence
on the momentum magnitude include
the momentum dependence of observables,
the possibility of increasing the CPT reach
by changing the boost of the mesons studied,
and even the possibility of increasing sensitivity
by a restriction to a subset of the data 
limited to a portion of the meson-momentum spectrum.
Consequences of the variation with 
momentum orientation include 
a dependence on the direction of the beam for collimated mesons,
a dependence on the angular distribution for other situations,
and diurnal effects arising from the rotation of the Earth
relative to the constant vector $\De\vec a$.

Since in real experiments 
the momentum and angular dependence
are often used experimentally to establish detector 
properties and systematics,
particular care is required to avoid subtracting 
or averaging away CPT-violating effects.
However,
observation of signals with momentum dependence
would represent a striking effect
that could help establish the existence
of CPT violation.
It can also suggest new ways of analyzing data
to increase the sensitivity of tests.
For example,
data taken with time stamps can be binned
according to sidereal (not solar) time
and used to constrain possible time variations
of observables as the Earth rotates
\cite{ak}.

The previous section established
the momentum dependence of various observables.
From the experimental perspective,
the expressions obtained can be regarded as
defined in the laboratory frame.
To display explicitly the time dependence
arising from the rotation of the Earth,
it is useful to exhibit the relevant expressions
in terms of parameters for CPT violation
expressed in a nonrotating frame.
In what follows,
the notation and conventions of Ref.\ \cite{kl}
are adopted,
with the spatial basis in the nonrotating frame denoted
$(\X,\Y,\Z)$
and that in the laboratory frame denoted 
$(\x,\y,\z)$.

In this coordinate choice, 
the basis $(\X,\Y,\Z)$ for the nonrotating frame 
is defined in terms 
of celestial equatorial coordinates
\cite{celestial}.
The rotation axis of the Earth defines the $\Z$ axis,
while $\X$ has declination and right ascension 0$^\circ$
and $\Y$ has declination 0$^\circ$ and
right ascension $90^\circ$.
This right-handed orthonormal basis
is independent of any particular experiment.
It can be regarded as constant in time
because the Earth's precession can be neglected
on the time scale of most experiments,
although care might be required in comparing
results between experiments performed
at times separated by several years or by decades. 

In the laboratory frame,
the most convenient choice of $\z$ axis 
depends on the experiment
but typically is along the beam direction.
For example,
if the experiment involves a collimated beam of mesons
the $\z$ direction can be taken as the direction of the beam.
If instead the experiment 
involves detecting mesons produced in a symmetric collider,
the $\z$ direction can be taken along 
the direction of the colliding beams at the intersection point.
Since time-varying signals are absent or reduced
if $\z$ is aligned with $\Z$,
in what follows these two unit vectors are taken to be different.
Then,
$\z$ precesses about $\Z$ with 
the Earth's sidereal frequency $\Om$,
and the angle $\ch\in(0,\pi)$ between the two unit vectors 
given by $\cos{\ch}=\z\cdot\Z$ is nonzero.
For definiteness,
choose
the origin of time $t=0$ such that $\z(t=0)$
lies in the first quadrant of the $\X$-$\Z$ plane.
Also, 
require $\x$ to be perpendicular to $\z$ 
and to lie in the $\z$-$\Z$ plane for all $t$:
$\x:=\z \cot\ch - \Z\csc\ch$.
Completing a right-handed orthonormal basis
with $\y:=\z\times\x$
means that 
$\y$ moves in the plane of the Earth's equator
and so is always perpendicular to $\Z$. 

Figure 1 shows the relation between the
two sets of basis vectors.
For ease of visualization only,
the basis $(\x,\y,\z)$ has been translated
from the laboratory location 
to the center of the globe.
Note, however, that 
at the location of the laboratory
$\z$ may lie at a non-normal angle to the Earth's surface.
Similarly,
the angle $\ch$ is unrelated to the colatitude of the experiment
unless the $\z$ axis happens to be normal to the Earth's surface
in the laboratory.

\begin{figure}
\centerline{\psfig{figure=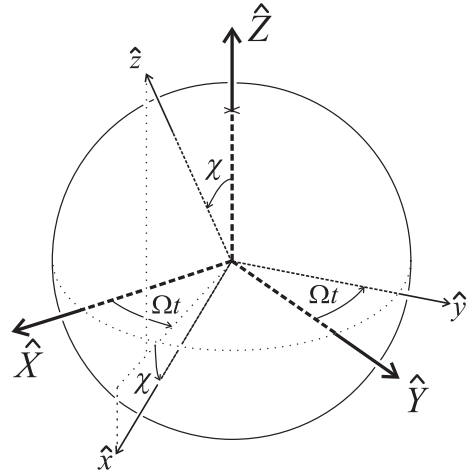,width=0.7\hsize}}
\caption{Bases in the laboratory and nonrotating frames.}
\label{fig:Figure1}
\end{figure}

Conversion between the two bases 
can be implemented with a nonrelativistic transformation,
given by Eq.\ (16) of Ref.\ \cite{kl}.
This assumes relativistic effects
due to the rotation of the Earth can be disregarded,
an assumption valid to about
one part in $10^{6}$ on the Earth's equator.
The time variation of
a parameter $\vec a \equiv (a^1, a^2, a^3)$
for Lorentz violation can then be directly obtained
in terms of its nonrotating-frame components
$(a^X, a^Y, a^Z)$:
\bea
a_1 (t) &=& 
a_X \cos\ch \cos\Om t 
+ a_Y \cos\ch \sin \Om t 
- a_Z \sin \ch
\quad ,
\nonumber\\
a_2 (t) &=& 
- a_X \sin\Om t 
+ a_Y \cos \Om t
\quad ,
\nonumber\\
a_3 (t) &=& 
a_X \sin\ch \cos\Om t 
+ a_Y \sin\ch \sin \Om t 
+ a_Z \cos \ch .
\label{at}
\eea
The above expressions permit
the time variation of the quantity $\vec \be \cdot \De \vec a$
and hence the time variation of various CPT observables
to be extracted.

The explicit form of the momentum and time dependence
of the parameter $\de_K$ in the kaon system 
can be found in the general case of a kaon with three-velocity
$\vec\be = \be (\sin\th\cos\ph, \sin\th\sin\ph, \cos\th)$
in the laboratory.
Here,
$\th$ and $\ph$ are conventional polar coordinates
defined in the laboratory frame about the $\z$ axis.
If the $\z$ axis is the beam axis,
these polar coordinates can be identified with 
the usual polar coordinates for a detector.
In terms of these quantities,
the above expressions yield 
\widetext
\top{-2.8cm}
\hglue -1 cm
\bea
\de_K(\vec p, t) &=& 
\fr {i \sin\hat\ph ~ e^{i\hat\ph}}
{\De m} \ga(\vec p)
\left[
\De a_0 + \be (\vec p) \De a_Z 
(\cos\th\cos\ch - \sin\th \cos\ph\sin\ch)
\right . 
\nonumber\\
&&
\qquad\qquad\qquad\qquad
\left. 
+\be (\vec p) \left(
-\De a_X \sin\th\sin\ph 
+\De a_Y (\cos\th\sin\ch + \sin\th\cos\ph\cos\ch )
\right) \sin\Om t
\right . 
\nonumber\\
&&
\qquad\qquad\qquad\qquad
\left. 
+\be (\vec p) \left(
\De a_X (\cos\th\sin\ch + \sin\th\cos\ph\cos\ch )
+\De a_Y \sin\th\sin\ph 
\right) \cos\Om t
\right]
\quad ,
\label{dept}
\eea
\bottom{-2.7cm}
\narrowtext
\noindent
where 
$\ga(\vec p) = \sqrt{1 + |\vec p|^2/m_K^2}$
and $\be(\vec p) = |\vec p|/m\ga(\vec p)$,
as usual.

The expression \rf{dept} is directly relevant
for specific experimental and theoretical analyses,
including those in the following subsections.
One feature of this expression is that
the complex phase of $\de_K$ is 
determined by $i \exp (i\hat\ph)$,
which is independent of momentum and time.
The real and imaginary parts of $\de_K$
therefore exhibit the same momentum 
and time dependence.
For instance,
$\Re\de_K$ and $\Im\de_K$
scale together if a meson is boosted.
Another feature is that the nature of the CPT-violating effects 
experienced by a meson varies with its boost.
For instance, 
if $\De \vec a = 0$ in the laboratory frame
then a boosted meson experiences a CPT-violating
effect greater by the boost factor $\ga$ 
relative to a meson at rest.
In contrast,
if $\De a_0 = 0$ in the laboratory frame
then there is no CPT-violating effect 
for a meson at rest
but there can be effects for a boosted meson.
The angular dependence in Eq.\ \rf{dept}
plays an important role in the latter case.
The variation of the size of $\de_K$ according
to sidereal time $t$
adds further complications,
including the possibility of effects averaging to zero
if, as usual, data are taken over extended time periods
and no time analysis is performed.

Evidently,
the momentum and time dependence displayed in Eq.\ \rf{dept}
implies that details of the experimental setup
play a crucial role in the analysis
of data for CPT-violating effects.
Next,
some issues relevant to two specific
and substantially different experiments are discussed.
Subsection \ref{fermilab} examines some aspects 
of an experiment involving collimated uncorrelated kaons 
with a nontrivial momentum spectrum and high mean boost.
Subsection \ref{daphne} considers
a few issues for an experiment producing correlated kaon pairs 
from $\ph$ decay in a symmetric collider.
The discussions are intended as illustrative
rather than comprehensive,
since a complete treatment of each case  
would require detailed simulation of the experimental conditions
and lies beyond the scope of this work.
The examples provided are chosen to improve intuition 
about the disparate consequences of 
the momentum and time dependence 
of the type \rf{dept} in CPT observables 
and to motivate their careful experimental study.

\subsection{Boosted Uncorrelated Kaons}
\label{fermilab}

In this subsection,
a few implications of the momentum and time dependence
are considered for a particular experimental scenario
involving CPT tests with boosted uncorrelated kaons.
Among possible experiments of this type
is the KTeV experiment presently underway at FNAL 
\cite{ktev}.
The kaon beam in this experiment is highly collimated
and has a momentum spectrum with an average boost 
factor $\overline\ga$ of order 100,
so $\be\simeq 1$.
The geometry is such that 
$\z\cdot\Z = \cos\ch \simeq 0.6$.

In experiments of this type,
the expression \rf{dept} for the momentum and time
dependence of $\de_K$ simplifies 
because the kaon three-velocity reduces to 
$\vec\be = (0,0,\be )$ in the laboratory frame. 
One finds
\bea
\de_K (\vec p, t) &=& 
\fr {i \sin\hat\ph ~ e^{i\hat\ph}} {\De m} \ga
[ \De a_0 + \be \De a_Z \cos\ch 
\nonumber\\
&& + \be \sin\ch ( \De a_Y \sin\Om t + \De a_X \cos\Om t ) ] .
\label{deptktev}
\eea
All four terms in this expression
depend on momentum through the relativistic factor $\ga$.
The first two exhibit no time dependence,
while the last two oscillate about zero with 
the Earth's sidereal frequency $\Om$.
Note that a conventional analysis 
seeking to constrain the magnitude $|\de_K|$ 
while disregarding the momentum and time dependence
would typically be sensitive to the average value
\beq
|\overline {\de_K}| = 
\fr {\sin\hat\ph } {\De m} \overline{\ga}
| \De a_0 + \overline{\be} \De a_Z \cos\ch |
\quad ,
\label{deptktevav2}
\eeq
where 
$\overline{\be}$ and $\overline{\ga}$ 
are the weighted averages of $\be$ and $\ga$,
respectively,
taken over the momentum spectrum of the data.

In many experiments, 
including KTeV,
$\de_K$ is indirectly reconstructed from other observables.
It is therefore of interest to identify the
momentum and time dependence of 
the quantities measured experimentally.
These include, for example,
the mass difference $\De m$,
the $K_S$ lifetime $\ta_S= 1/\ga_S$,
and the ratios $\et_{+-}$, $\et_{00}$ 
of amplitudes for $2\pi$ decays,
defined as usual by
\bea 
\et_{+-} &\equiv &
\fr {A(K_L \to \pi^+\pi^-)} {A(K_S \to \pi^+\pi^-)}
\equiv | \et_{+-} | e^{i\ph_{+-}} 
\approx \ep + \ep^\prime
\quad ,
\nonumber \\
\et_{00} &\equiv &
\fr {A(K_L \to \pi^0\pi^0)} {A(K_S \to \pi^0\pi^0)}
\equiv | \et_{00} | e^{i\ph_{00}}
\approx \ep - 2\ep^\prime
\quad .
\label{etas}
\eea
In the Wu-Yang phase convention \cite{wy},
it can be shown that
$\ep \approx \ep_K - \de_K$ 
\cite{barmin,td}.
Note that $|\ep| \simeq 2\times 10^{-3}$
\cite{pdg}
and that $|\ep^\prime| \simeq 6\times 10^{-6}$
\cite{ktevcp}.

Consider first the case where $|\ep_K| > |\de_K| > |\ep^\prime|$.
This corresponds to the current experimental situation,
since $|\de_K|$ is presently constrained to about $10^{-4}$.
Neglecting $\ep^\prime$ then gives
\bea
| \et_{+-} | e^{i\ph_{+-}}
&\approx &
| \et_{00} | e^{i\ph_{00}} 
\approx 
\ep \approx \ep_K - \de_K 
\nonumber\\
&\approx & (|\ep_K| + i |\de_K|)e^{i\hat\ph}
\quad ,
\label{approx}
\eea
where the last expression follows because 
the phases of $\ep_K$ and $\de_K$ differ by $90^\circ$
\cite{buch}.
Then,
it follows that 
\bea
|\et_{+-}| & \approx & 
|\et_{00}|  \approx 
|\ep_K|(1 + O(|\de_K/\ep_K|^2)
\quad ,
\nonumber\\
\ph_{+-} &\approx & 
\ph_{00} \approx 
\hat\ph + |\de_K/\ep_K|
\quad .
\label{approx2}
\eea

The above expressions
and the results in the previous section
show that at leading order in CPT-violating parameters
the only observable quantities 
exhibiting leading-order momentum and time dependence 
are the phases $\ph_{+-}$ and $\ph_{00}$.
In terms of experimental observables 
and parameters for CPT violation,
one finds
\bea
\ph_{+-} &\approx & \ph_{00} 
\nonumber\\
&\approx &\hat\ph + 
\fr {\sin\hat\ph } {|\et_{+-}| \De m} \ga
[ \De a_0 + \be \De a_Z \cos\ch 
\nonumber\\
&& 
\qquad
+ \be \sin\ch ( \De a_Y \sin\Om t + \De a_X \cos\Om t) ] .
\label{deptktev3}
\eea
The other experimental observables
$|\et_{+-}|$, $|\et_{00}|$, $\ep^\prime$,
$\De m$, $\hat\ph$, $\ta_S= 1/\ga_S$
either have no momentum and time dependence
or have it suppressed by the square of the
parameters for CPT violation.

The result \rf{deptktev3} shows that an experiment
using collimated kaons with a momentum spectrum
can in principle place independent bounds
on each of the components 
$\De a_0$, $\De a_X$, $\De a_Y$, $\De a_Z$.
The variation with sidereal time 
in Eq.\ \rf{deptktev3} is a sum of sine and cosine terms,
equivalent to a pure sinusoidal variation
of the form $A_{+-}\sin(\Om t + \al)$
with amplitude $A_{+-}$ proportional to
$a_\perp\equiv \sqrt{(\De a_X)^2 + (\De a_Y)^2}$
and phase $\al$ determined by the ratio $\De a_X/\De a_Y$:
\beq
A_{+-} = 
\be\ga\fr {\sin\hat\ph \sin\ch} {|\et_{+-}| \De m} a_\perp , \quad 
\tan\al = \De a_X/\De a_Y
\quad .
\label{amp}
\eeq
Binning in time would in principle allow independent
constraints on $\De a_X$ and $\De a_Y$.
A time-averaged analysis would provide a bound on 
the combination $\De a_0 + \be \De a_Z \cos\ch$,
from which $\De a_0$ and $\De a_Z$ could be 
separated if the momentum spectrum is sufficiently broad
to permit useful binning in $\be$.
In the specific case of the KTeV experiment,
it may be feasible to disentangle $\De a_X$ and $\De a_Y$,
but separation of $\De a_0$ and $\De a_Z$ 
may be difficult to achieve.

Although the derivation of Eq.\ \rf{deptktev3} neglects  
the role of $\ep^\prime$,
the small size of the latter ensures that
the results remain valid to leading order
in CPT-violating quantities
provided $|\de_K|$ is not too small.
Since the observable CPT violation is predicted to occur
only in the mixing matrix,
$\ep^\prime$ itself is not directly affected.
This restricts its role to momentum- and time-independent
corrections to the above equations
or to suppressed contributions controlled
by the product $\ep^\prime\de_K$.
For example,
if $\ep^\prime$ is included one finds 
Eq.\ \rf{deptktev3} acquires a correction $- \De\ph/3$:
\bea
\ph_{+-} 
&\approx &\hat\ph - \frac 1 3 \De\ph 
\nonumber\\
&&+\fr {\sin\hat\ph } {|\et_{+-}| \De m} \ga
[ \De a_0 + \be \De a_Z \cos\ch 
\nonumber\\
&& 
\qquad
+ \be \sin\ch ( \De a_Y \sin\Om t + \De a_X \cos\Om t) ] ,
\label{modified}
\eea
where $\De\ph \equiv \ph_{00} - \ph_{+-}$.
However,
this difference contains at most higher-order 
CPT-violating effects 
\cite{fn3}.

The occurrence of momentum and time dependence in the
observables for CPT violation
has a variety of consequences
for the analysis of data 
in experiments of the type considered in this subsection
\cite{ak}.
These range from direct implications 
such as distinctive CPT signals
correlated with the momentum spectrum and time stamps
to more indirect consequences 
such as a CPT reach proportional to the boost factor 
$\ga$ under suitable circumstances.
There are also some more exotic implications,
such as the possibility in principle of 
improving the CPT reach under certain circumstances
by examining only a specific momentum range of a subset 
of the available data.
Evidently,
a careful experimental analysis allowing
for the effects of possible momentum and time dependence
has the potential to yield interesting results. 

\subsection{Unboosted Correlated Kaon Pairs}
\label{daphne}

In this subsection,
some effects of the momentum and time dependence
are considered in an experiment 
testing CPT using correlated kaon pairs 
from $\ph$ decay in a symmetric collider.
The KLOE experiment 
\cite{kloe}
at DAPHNE in Frascati provides an example of this kind.
These experimental circumstances differ substantially 
from those of the example in the previous subsection,
and as a result the CPT reach is controlled by different factors.
The origin of the kaon pairs
in the decay from the $\ph$ quarkonium state just above threshold
implies a line spectrum in the laboratory-frame momentum 
of about 0.1 GeV for each kaon, 
so the momentum dependence of the CPT observables 
is relatively uninteresting.
In contrast,
significant implications for the experiment
arise from the wide angular distribution of  
the kaon momenta in the laboratory frame.

Consider first the general situation
of a quarkonium state with $J^{PC}=1^{--}$
decaying at time $t$ in its rest frame
into a correlated $P$-$\overline{P}$ pair.
Since the laboratory frame coincides with 
the quarkonium rest frame,
the time $t$ can be identified with the
sidereal time introduced earlier.
For $P\equiv K$
the relevant quarkonium is the $\ph$ meson,
while for $P\equiv B_d$ it is $\Up (4S)$,
for $P\equiv B_s$ it is $\Up (5S)$,
and for $P\equiv D$ it is $\ps (3770)$.

Immediately following the strong decay of the quarkonium,
the normalized initial state $\ket i$ has the form
\cite{lipkin}
\beq
\ket i = N \left[
\ket{P_S(+)P_L(-)}
-\ket{P_L(+)P_S(-)}
\right]
\quad ,
\label{iva}
\eeq
where $N$ is a normalization.
In this expression,
the eigenstates of the effective hamiltonian are denoted by
$\ket{P_S (\pm)}$ and $\ket{P_L (\pm)}$,
where $(+)$ means the particle is moving in a specified direction 
and $(-)$ means it is moving in the opposite direction.

Suppose at time $t + t_1$ in the quarkonium rest frame 
the meson moving 
in the $(+)$ direction decays into $\ket{f_1}$,
while the other decays into $\ket{f_2}$ at $t+t_2$.
Define for each $\al$ the ratio of amplitudes
\beq
\et_\al
\equiv|\et_\al|e^{i\ph_\al}
= \fr {A(P_L\to f_\al) }{A(P_S\to f_\al)}
\quad .
\label{ivc}
\eeq
Note that some of these quantities may depend 
on the momentum and time through a possible dependence on $\De M$.
Then,
the net amplitude 
${\cal{A}}_{12}(\vec p, t, t_1, t_2)$ for the 
correlated double-meson decay into $f_1$ and $f_2$ is
\widetext
\top{-2.8cm}
\hglue -1 cm
\beq
{\cal{A}}_{12}(\vec p, t ,t_1, t_2)
=\hat N 
\bigl (
 \eta_2e^{-i(m_St_1+m_Lt_2)-\half (\ga_St_1+\ga_Lt_2)}
-\eta_1e^{-i(m_Lt_1+m_St_2)-\half (\ga_Lt_1+\ga_St_2)}
\bigr ) 
\quad ,
\label{ivd}
\eeq
where $\hat N = N {A(P_S\to f_1) }{A(P_S\to f_2)}$.
In this expression,
the possible dependence 
on the three-momenta 
$\vec p_1 = - \vec p_2 \equiv \vec p$
of the two mesons and on the sidereal time $t$
has been suppressed
in the right-hand side of Eq.\ \rf{ivd},
but if present it would reside in $\et_\al$ and $\hat N$.

Taking the modulus squared of the amplitude \rf{ivd} 
gives the double-decay rate.
In terms of 
$\overline{t} = t_1+t_2$
and
$\De t = t_2 - t_1$,
the double-decay rate 
$R_{12}(\vec p,t,\overline{t},\De t)$ 
is 
\beq
R_{12}(\vec p,t,\overline{t},\De t)
= |\hat N|^2 
 e^{- \overline{\ga} \overline{t}/2}
\left [
|\eta_1|^2 e^{- \De \ga \De t/2}
+|\et_2|^2 e^{\De \ga \De t/2}
- 2|\et_1\et_2|\cos(\De m\De t + \De \ph)
\right ] ,
\label{ivg}
\eeq
\bottom{-2.7cm}
\narrowtext
\noindent
where $\overline{\ga} = \ga_S+\ga_L$ and
$\De \ph = \ph_1 - \ph_2$.

A detailed study of the CPT signals from 
symmetric-collider experiments with correlated mesons 
requires simulation with expressions
of the type \rf{ivg} for various final states $f_1$, $f_2$. 
Given sufficient experimental resolution,
the dependence of certain decays on the two meson momenta 
$\vec p_1$, $\vec p_2$ and on the time $t$
could be exhibited experimentally 
by appropriate data binning and analysis.
However,
some caution is required because 
different asymmetries can be sensitive to 
distinct components of $\De M$.

Consider,
for example,
the case of double-semileptonic decays 
of correlated kaon pairs
in a symmetric collider.
Neglecting violations of the $\De S = \De Q$ rule,
for the state $f_+ \equiv l^+ \pi^- \nu$ 
one finds $\et_{l^+} \approx 1 - 2 \de_K$,
while for 
$f_- \equiv l^- \pi^+ \overline{\nu}$
one finds $\et_{l^-} \approx -1 - 2 \de_K$.
Inspection of Eq.\ \rf{ivg} shows 
that the double-decay rate $R_{l^+l^-}$ 
can be regarded as proportional to an expression
depending on the ratio
\bea
\left| \fr{\et_{l^+}}{\et_{l^-}}\right| 
&\approx &
1 - 2 \Re\de_K(+) - 2 \Re\de_K(-)
\nonumber\\
&=&
1 - \fr {4\Re (i \sin\hat\ph ~ e^{i\hat\ph})} 
{\De m} \ga(\vec p) \De a_0 
\quad .
\label{r+-}
\eea
In the first line of the above expression,
the CPT-violating contributions 
from each of the two kaons have been kept distinct
because they differ in general 
for mesons traveling in different directions.
All the angular and time dependence in Eq.\ \rf{dept}
cancels from the second line because 
in the present case of a symmetric collider
$\vec\be_1\cdot\De\vec a = - \vec\be_2\cdot\De\vec a$.

The proportionality factor
for the double-decay rate $R_{l^+l^-}$
is $|\hat N\et_{l^-}|^2$,
which depends on the full expression \rf{dept} for $\de_K$. 
However,
this factor plays the role of a normalization.
Unless it is carefully tracked,
which could be a potentially challenging experimental task,
no angular or time dependence would be manifest
in the double-semileptonic decay mode.
For instance,
the normalization factor would play no role 
in a conventional analysis 
to extract the physics using an asymmetry,
for which normalizations cancel.
Moreover,
as mentioned above,
the line spectrum in the momentum means that
the dependence on $|\vec p|$ is also unobservable. 
Indeed, $\ga(\vec p) \simeq 1$.
The double-semileptonic decay is therefore
well suited to placing a clean constraint on 
the timelike parameter $\De a_0$ for CPT violation,
and the experimental data can be collected for analysis 
without regard to their angular locations in the detector
or their sidereal time stamps.

In contrast,
certain mixed double-decay modes
are sensitive to $\de_K$ only in one of the two decays.
This is the case,
for instance,
for double-decay modes
with one semileptonic prong and one double-pion prong.
In these situations,
the double-decay rate $R_{12}$ in Eq.\ \rf{ivg} 
can be directly sensitive to the angular and
time dependence exhibited in Eq.\ \rf{dept},
and in particular it can provide sensitivity
to the parameters $\De\vec a$ for CPT violation.
In a conventional analysis,
CPT violation in a given double-decay mode of this type
is inextricably linked with other parameters for CP violation
\cite{buch,rosner,hs}.
However,
in the present case
the possibility of binning for angular and time dependence
means that clean tests of CPT violation are feasible
even for these mixed modes.

Consider, 
for example,
a detector with acceptance independent of the azimuthal angle $\ph$.
The distribution of mesons from the quarkonium decay 
is symmetric in $\ph$,
so the $\de_K$-dependence of a $\ph$-averaged dataset 
is governed by the expression
\bea
\de_K^{\rm av}(|\vec p|, \th, t) 
&\equiv &\fr 1 {2\pi} \int_0^{2\pi} d\ph ~\de_K(\vec p, t) 
\nonumber\\
&=& \fr {i \sin\hat\ph ~ e^{i\hat\ph}}
{\De m} \ga
\left[
\De a_0 + \be \De a_Z \cos\ch \cos\th
\right . 
\nonumber\\
&&
\quad \qquad
\left. 
+\be \De a_Y \sin\ch \cos\th \sin\Om t 
\right . 
\nonumber\\
&&
\quad \qquad
\left. 
+\be \De a_X \sin\ch \cos\th \cos\Om t
\right] 
\quad .
\label{deptph}
\eea
Inspection shows that by binning in $\th$ and in $t$
an experiment with asymmetric double-decay modes 
can in principle extract separate bounds 
on each of the three components
of the parameter $\De \vec a$ for CPT violation.
This result holds independent of other CP parameters 
that may appear,
since the latter have neither angular nor time dependence.
Combining data from asymmetric double-decay modes
and from double-semileptonic modes 
therefore permits in principle
the extraction of independent bounds 
on each of the four components of $\De a_\mu$.

The same reasoning applies to other 
experimental observables.
For example,
one can consider the standard rate asymmetry
for $K_L$ semileptonic decays \cite{pdg},
\bea
\de_l &\equiv &
\fr{\Ga (K_L \to l^+\pi^-\nu) 
- \Ga(K_L \to l^-\pi^+\overline{\nu})}
{\Ga (K_L \to l^+\pi^-\nu) 
+ \Ga(K_L \to l^-\pi^+\overline{\nu})}
\nonumber \\
&&\approx 2\Re\ep_K - 2\Re\de_K (\vec p, t)
\quad ,
\label{dell}
\eea
where the symbol $\Ga$ denotes a partial decay rate
and $\De S = \De Q$ has been assumed.
In principle,
this asymmetry could also be studied for angular
and time variation,
leading to constraints on $\De a_\mu$. 

\section{Summary}
\label{summary}

This paper has considered some issues
involving momentum- and time-dependent
experimental signals for indirect CPT violation
in a neutral-meson system.
Effects of this type are predicted
in the context of a general standard-model extension
allowing for CPT and Lorentz violation.
Their consequences for data analysis vary substantially
with the specifics of a given experiment.

Following a theoretical description of the 
momentum and time dependence 
applicable to any neutral-meson system,
specific theoretical results are obtained for kaons.
Some CPT observables depend explicitly on 
the magnitude and orientation of the meson momentum,
which leads to diurnal variations in observables.

Illustrations of the consequences of these results
for experiments are provided 
using two types of scenario already adopted
at FNAL and Frascati,
one with collimated boosted uncorrelated kaons
and the other with uncollimated correlated kaon pairs
from $\ph$ decay at rest.
The implications described for these scenarios
also provide intuition about
possible effects in other experiments with kaons
and in experiments with $D$ or $B$ mesons.

The primary message of this work is that 
a complete extraction of CPT bounds from any experiment 
allowing for possible momentum dependence and diurnal variation
of the observables is a worthwhile undertaking
and one that has the potential to yield further surprises
from the enigmatic neutral-meson systems. 

\section*{Note added in press}
The KTeV collaboration has reported
\cite{k99}
a preliminary bound $A_{+-} \lsim  0.5 ^\circ$
on the amplitude \rf{amp} of variations 
of $\ph_{+-}$ with sidereal time,
corresponding to a constraint
$a_\perp \lsim 10^{-20}$ GeV.

\section*{Acknowledgments}
This work is supported in part 
by the Department of Energy
under grant number DE-FG02-91ER40661.

\end{multicols}

\begin{references}
\bibitem{qmviol}
It has been suggested 
that in the kaon system
CPT-violating effects might be generated
through an unconventional quantum mechanics 
in which the Schr\"odinger equation
is replaced with a density-matrix formalism.
See J.\ Ellis {\it et al.},
Phys.\ Rev.\ D {\bf 53} (1996) 3846,
where it is also shown that the resulting effects
can be disentangled from the CPT-violating
parameter $\de_K$ of interest here.
\bibitem{fn0}
Direct CPT violation in the decay amplitudes
is neglected here because it is suppressed
in the standard-model extension
and is expected to be unobservable
\cite{kp}.
\bibitem{fn1}
A field redefinition 
can be used to shift equally the values
of all the parameters $a_\mu^q$
without physical consequences 
\cite{cksm}.
For a particular $P^0$ meson,
one could therefore replace $\De a_\mu$ 
with a single $a_\mu$ factor.
For clarity,
this redefinition is avoided in the present work.
\end{references}
\end{document}